\documentclass{aa}
\usepackage{graphics,epsfig,lscape,times}

\usepackage{natbib}
\bibpunct{(}{)}{;}{a}{}{,}

\newcommand{\oergs}[1]{$10^{#1}$ erg s$^{-1}$}

\newcommand{\expo}[1]{$\times 10^{#1}$}
\newcommand{\oexpo}[1]{$10^{#1}$}
\newcommand{\nh}{N$_{\rm H}$}
\newcommand{\gr}{\hbox{g$_{\rm r}$}}
\newcommand{\ecp}{\hbox{E$_{\rm cp}$}}
\newcommand{\eline}{\hbox{E$_{\rm line}$}}

\newcommand{\eqw}{\hbox{EQW}}

\newcommand{\ltsima}{$\buildrel < \over \sim$}
\newcommand{\lsim}{\lower.5ex\hbox{\ltsima}}
\newcommand{\gtsima}{$\buildrel > \over \sim$}
\newcommand{\gsim}{\lower.5ex\hbox{\gtsima}}
\newcommand{\rbs}{\hbox{\object{RBS1223}}}
\newcommand{\rxj}{\hbox{\object{1RXS\,J130848.6+212708}}}
\newcommand{\rxa}{\hbox{\object{RX\,J1856.4$-$3754}}}
\newcommand{\rxb}{\hbox{\object{RX\,J0720.4$-$3125}}}
\begin{document}
 
\title{A broad absorption feature in the X-ray spectrum of the isolated neutron star
       RBS1223 (1RXS\,J130848.6+212708)
        \thanks{Based on observations with XMM-Newton,
               an ESA Science Mission with instruments and contributions 
	       directly funded by ESA Member states and the USA (NASA)}}
 
\author{F.~Haberl\inst{1} \and A.D.~Schwope\inst{2} \and V.~Hambaryan\inst{2} \and
        G.~Hasinger\inst{1} \and C.~Motch\inst{3}}

\titlerunning{An absorption line in the X-ray spectrum of RBS1223}
\authorrunning{Haberl et al.}
 
\offprints{F. Haberl, \email{fwh@mpe.mpg.de}}
 
\institute{Max-Planck-Institut f\"ur extraterrestrische Physik,
           Giessenbachstra{\ss}e, 85748 Garching, Germany
	   \and
	   Astrophysikalisches Institut Potsdam, An der Sternwarte 16,
	   14482 Potsdam, Germany
	   \and
	   Observatoire Astronomique, UA 1280 CNRS, 11 rue de l'Universit{\'e}, 67000
	   Strasbourg, France}
 
\date{Received 6 March 2003 / Accepted 25 March 2003}
 
\abstract{
X-ray spectra of the isolated neutron star RBS1223 obtained with the
instruments on board XMM-Newton in December 2001 and January 2003
show deviations from a Planckian energy
distribution at energies below 500 eV. The spectra are well fit when a broad, 
Gaussian-shaped absorption line with $\sigma$ = 100 eV and centered at an 
energy of 300 eV is added to an absorbed blackbody model. The resulting
equivalent width of the line is $-$150 eV. However, the spectral resolution at these 
low energies of the EPIC detectors and the lower statistical quality and restricted
energy band of the RGS instruments are not sufficient to exclude even broader lines 
at energies down to 100 eV or several unresolved lines. The most likely interpretation 
of the absorption feature is a cyclotron absorption line produced by
protons in the magnetic field of the neutron star. In this picture line energies
of 100$-$300 eV yield a magnetic field strength of 2$-$6\expo{13} G for a neutron star
with canonical mass and radius. Folding light curves from different 
energy bands at a period of 10.31 s, which implies a double peaked pulse profile,
shows different hardness ratios for the two peaks. This confirms that the true spin
period of \rbs\ is twice as long as originally thought and suggests variations 
in cyclotron absorption with pulse phase. We also propose that changes in photo-electric 
absorption seen in phase resolved spectra of \rxb\ by \citet{2001A&A...365L.302C},
when formally fit with an absorbed blackbody model, are caused instead by cyclotron 
absorption varying with pulse phase.
}
 
\maketitle
 
\section{Introduction}

The soft X-ray source \rxj\ was discovered in the ROSAT all-sky survey data
and further studied as part of the ROSAT Bright Survey program (RBS, source number 1223) 
to optically identify the brightest $\sim$2000 high-galactic latitude 
sources \citep{2000AN....321....1S}. 
Based on similarities to the two best known cases
\rxa\ and \rxb\ and the lack of an optical counterpart brighter than B$\sim$26$^m$,
\citet{1999A&A...341L..51S} proposed \rxj\ = \rbs\ as isolated neutron star (INS). 
Using the Chandra observatory \citet{2002A&A...381...98H} observed
\rbs\ on June 24, 2000 with the ACIS instrument and discovered pulsations in the
X-ray flux, initially indicating a neutron star spin period of 5.16 s. 
A probable optical counterpart in the 90\%
Chandra error circle was reported by \citet{2002ApJ...579L..29K} using a very 
deep observation from the Hubble Space Telescope. The optical brightness of
\hbox{$m_{\rm 50CCD}$} = 28.6$^m$ yields an X-ray to optical flux ratio of 
log(f$_{\rm x}$/f$_{\rm opt}$) = 4.9. The optical flux is a factor of about 5 above the
extra\-po\-lation of the blackbody fit to the X-ray spectrum, similar as observed for 
\rxa\ \citep[e.g.][]{burwitz} and \rxb\ \citep{2002nsps.conf..273P}.

This confirms \rbs\ as member of a group of X-ray dim INSs which share 
similar properties \citep[for a recent review see][]{haberl2002COSPAR}: soft blackbody-like 
X-ray spectrum, radio quiet and no known association with a supernova remnant. Four 
objects are X-ray pulsars with neutron star spin periods in the range 8.39 to 22.7 s 
\citep[the XMM-Newton observations of \rbs\ revealed a peak in the power spectrum at
half the frequency of the main peak which suggests that the genuine spin period 
is 10.31 s with a double peaked profile,][]{haberl2002COSPAR}. 
All available spectra obtained by the ROSAT
PSPC were consistent with Planckian energy distributions with blackbody 
temperatures kT in the range 40 $-$ 100 eV and little attenuation by interstellar 
absorption. More recent observations of
the two brightest objects \rxa\ \citep{2001A&A...379L..35B,burwitz} and \rxb\ 
\citep{2001A&A...365L.298P,2002nsps.conf..273P} were performed using the low
energy transmission grating (LETG) aboard Chandra and the reflection grating 
spectrometers (RGS) of XMM-Newton. Also at high spectral resolution both sources 
show featureless spectra which can best be modeled by a Planckian spectrum with 
kT of $\sim$63 eV and $\sim$86 eV, respectively.

The pulsars among the X-ray dim INSs may allow an estimate of the strength of their
magnetic field if their long-term spin period increase can be interpreted in terms of 
dipolar losses. For the pulsar \rxb\ \citet{2002ApJ...570L..79K} derived 
an upper limit for the period derivative of 3.6\expo{-13} s s$^{-1}$ from the analysis 
of ROSAT, SAX and Chandra data and including first XMM-Newton data, \citet{2002MNRAS.334..345Z}
determined possible values between 3$-$6\expo{-14} s s$^{-1}$. This implies that the magnetic
field strength of \rxb\ could be as high as a few \oexpo{13} G.
For such magnetic field strengths an absorption feature at the proton cyclotron
resonance energy, \ecp\ = 63\gr(B/\oexpo{13} G) eV (\gr\ = 
(1$-$2GM/c$^2$R)$^{1/2}$ with M and R the neutron star mass and radius, respectively), 
may become detectable \citep{2001ApJ...560..384Z,2002nsps.conf..263Z} with  
detectors sensitive at soft X-rays down to 100 eV.

In this letter we present the results from a spectral analysis of two XMM-Newton 
observations of \rbs. The spectra show deviations from a Planckian energy distribution
which can best be modeled by a broad absorption feature at low energies. We 
discuss this feature as most likely being caused by proton cyclotron resonance 
absorption.

\section{XMM-Newton observations}

XMM-Newton \citep{2001A&A...365L...1J} observed \rbs\ on two occasions;
in the guest observer program (AO1, PI: Schwope) and for calibration purposes at 
the beginning of the year 2003.
Here we utilize the data collected with the European Photon Imaging Cameras 
(EPICs) and the Reflection Grating Spectrometers 
\citep[RGSs,][]{2001A&A...365L...7D} for a spectral 
analysis. Two EPIC cameras are
based on MOS \citep[EPIC-MOS1 and \hbox{-MOS2,}][]{2001A&A...365L..27T}
and one on pn \citep[EPIC-pn,][]{2001A&A...365L..18S} CCD detectors and
are mounted behind the three X-ray multi-mirror systems. 
The details of the XMM observations are summarized in Table~\ref{epic-obs}.

\begin{table}
\caption[]{XMM-Newton observations of \rbs.}
\begin{tabular}{cccccc}
\hline\noalign{\smallskip}
\multicolumn{2}{c}{Time (UT)} &
\multicolumn{1}{l}{XMM} &
\multicolumn{1}{c}{Read-out} &
\multicolumn{1}{c}{Count rate} &
\multicolumn{1}{c}{Exp.} \\
\multicolumn{1}{c}{Start} &
\multicolumn{1}{c}{End} &
\multicolumn{1}{l}{Detector} &
\multicolumn{1}{c}{Mode} &
\multicolumn{1}{c}{[s$^{-1}$]} &
\multicolumn{1}{c}{[ks]} \\

\noalign{\smallskip}\hline\noalign{\smallskip}
\multicolumn{6}{l}{2001 Dec. 31, Satellite Revolution 377 (AO1)} \\
\noalign{\smallskip}\hline\noalign{\smallskip}
 03:08 & 08:41 & MOS1/2  & Full Frame	& 0.49/0.52 & 19.8 \\
 03:32 & 08:42 & pn	 & Small Window &    2.47   & 19.0 \\
 03:02 & 08:42 & RGS1/2  & Spectro+Q	& 0.09/0.07 & 20.4 \\	
\noalign{\smallskip}\hline\noalign{\smallskip}
\multicolumn{6}{l}{2003 Jan. 1, Satellite Revolution 561 (CAL)} \\
\noalign{\smallskip}\hline\noalign{\smallskip}
 06:13 & 14:14 & MOS1/2  & Large Window & 0.47/0.48 & 28.7 \\
 06:36 & 14:14 & pn	 & Full Frame	&    2.39   & 27.0 \\
 06:12 & 14:15 & RGS1/2  & Spectro+Q	& 0.10/0.08 & 28.9 \\	
\noalign{\smallskip}
\hline
\end{tabular}

Mean count rates are given for the energy band used for the spectral analysis
(EPIC: 0.13$-$1.3 keV, RGS: 0.32$-$0.90 keV). The thin filter was used in all EPIC cameras
during both observations.
\label{epic-obs}
\end{table}

\section{X-ray spectra}

The data from the EPIC instruments were processed using SAS5.4.1 to produce
event files for pn, MOS1 and MOS2. RGS spectra and response files were created
using rgsproc of SAS5.4.1 with standard settings.
The EPIC spectra were extracted from circular regions with 30\arcsec\ radius
around the target position and nearby background regions using single-pixel events
(pattern 0) from pn and pattern 0$-$12 events from MOS data.
Detector response matrix files were used as available from the
EPIC-pn calibration group in February 2003 (version 6.5 which will
be implemented in the next SAS release after SAS5.4.1) 
and were created for MOS using the arfgen and rmfgen tasks of 
SAS5.4.1. MOS spectra from recent observations suffer from a considerable change in the
low energy re-distribution which is not calibrated yet. Therefore, we use MOS spectra 
from the first observation in Dec. 2001, only. The pn spectra from the two observations, 
obtained in different read-out modes, are in agreement within the current
calibration uncertainties and also are consistent with the MOS spectra from the Dec. 2001
observation.

The spectra were simultaneously fitted with various models, all including photo-electric
absorption with element abundances from \citet{1989GeCoA..53..197A} and cross-sections from
\citet{1992ApJ...400..699B} and \citet{1998ApJ...496.1044Y} using XSPEC version 11.2.0.
To account for cross-calibration uncertainties between the instruments and the fact that
the pn spectra were not corrected for point spread function losses (the SAS5.4.1 task arfgen 
is not compatible to the response files we used here) the relative normalization between the 
instruments/observations were allowed to vary individually. The pn fluxes were corrected 
manually by a factor 1.2 to correct for the PSF loss according to the XMM-Newton users 
handbook. In addition uncertainties in the 
low-energy re-distribution were accounted for by allowing individual column density values 
for pn, MOS1 and MOS2. The RGS spectra show relatively large differences in continuum
parameters which were then fit for each spectrum separately.

\begin{figure*}
\begin{center}
\hbox{
\resizebox{8.6cm}{!}{\includegraphics[clip=,angle=-90]{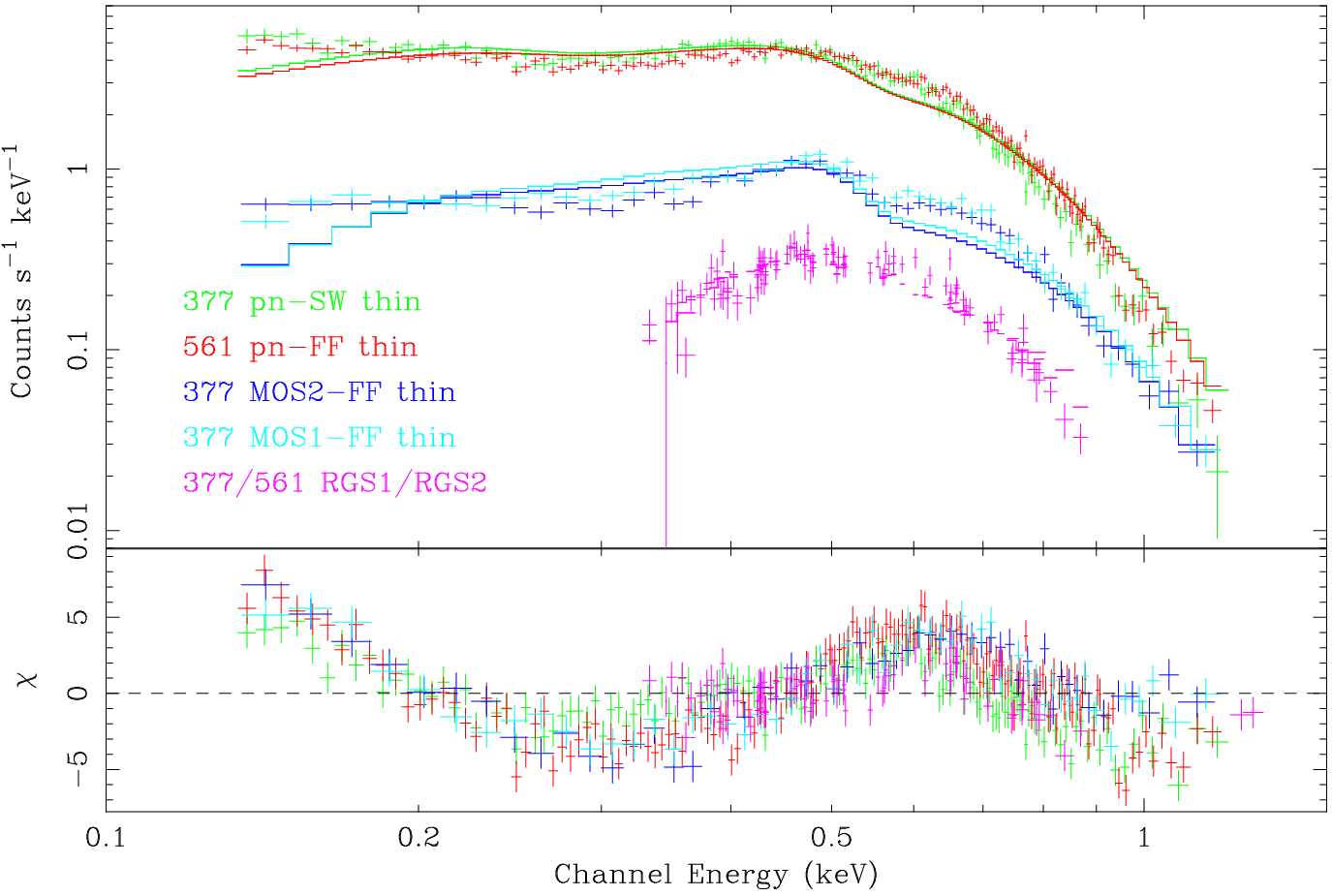}}
\hspace{4mm}
\resizebox{8.6cm}{!}{\includegraphics[clip=,angle=-90]{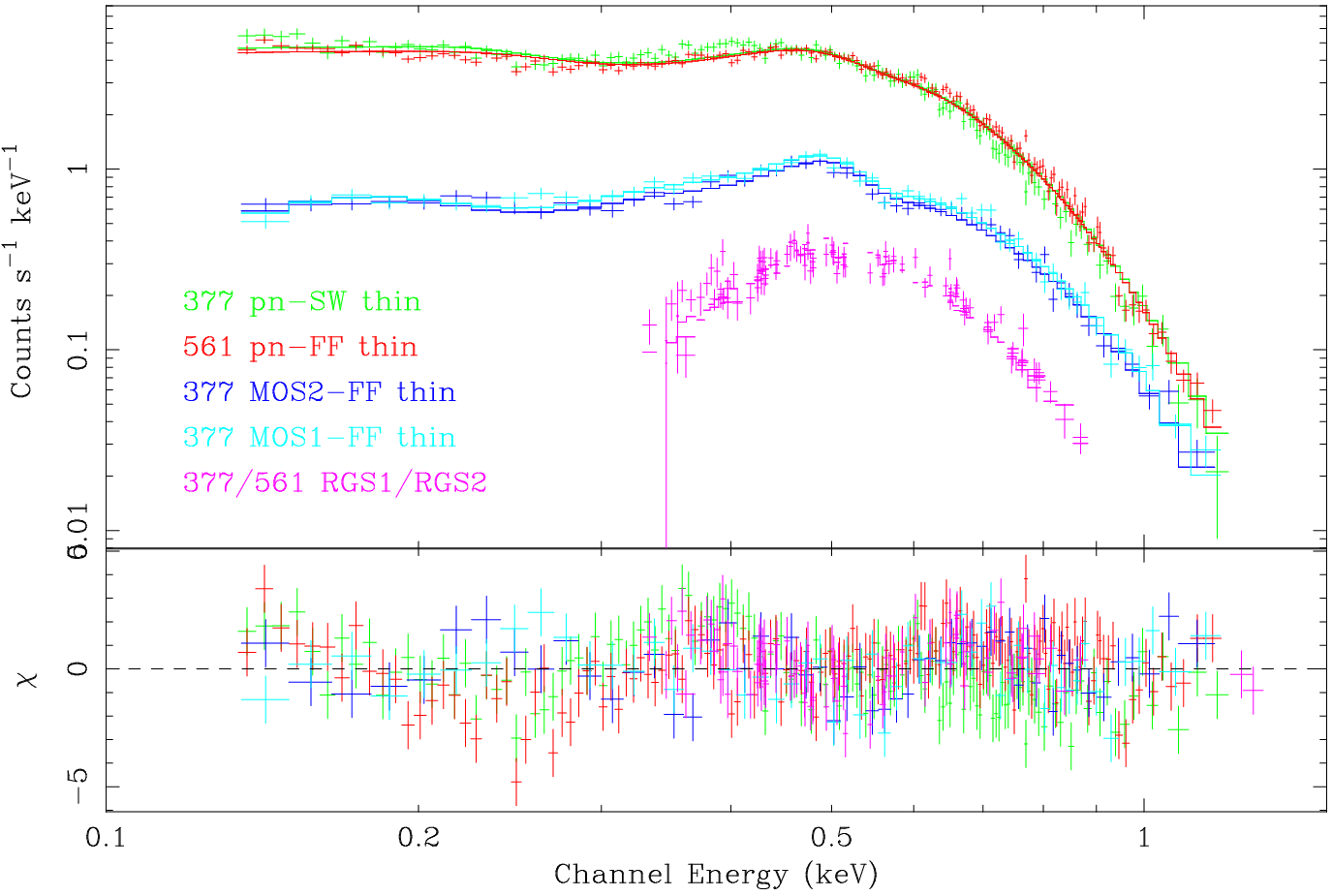}}
}
\end{center}
\caption{Blackbody model fits to EPIC-pn (upper pair), EPIC-MOS (middle pair) and
         RGS spectra of \rbs. The four RGS spectra were combined in the plot for 
	 clarity.
         While the pure blackbody model fit (left) is unacceptable, including a broad
         Gaussian absorption line at $\sim$300 eV (right) can reproduce the data.
	 The residuals (bottom panels) show consistent behavior for all instruments.}
\label{fig-spectra}
\end{figure*}

As first model an absorbed blackbody was used, however, the combined fit to the XMM spectra
of \rbs\ is not acceptable with $\chi^2_r$ of 5.2 for 639 degrees of freedom (dof).
The spectra together with the fit are shown in Fig.~\ref{fig-spectra}
and the fit parameters summarized in Table~\ref{tab-fits}. Because of the bad 
fit we do not give errors and fluxes for this model.
In particular, the EPIC spectra with best statistical quality yield high $\chi^2$ values
which we report for each instrument separately in Table~\ref{tab-fits}.
Since the combined fit is dominated by the high statistics of the EPIC-pn spectra we 
allowed all model parameters for the RGS spectra to be fit freely, i.e. the RGS fit
is completely independent from the pn fit. This allows to judge if the RGS spectra
on their own can be well fit by a blackbody model.

\begin{table*}[t]
\caption[]{Spectral fit results.}
\begin{tabular}{l|ccc|cccccc}
\hline\noalign{\smallskip}
\multicolumn{1}{l|}{} &
\multicolumn{3}{c|}{Model A: phabs*(bbody)} &
\multicolumn{6}{c}{Model B: phabs*(bbody+gaussian)} \\

\multicolumn{1}{l|}{Obs.-Instrument} &
\multicolumn{1}{c}{kT} &
\multicolumn{1}{c}{\nh} &
\multicolumn{1}{c|}{$\chi^2$/dof} &
\multicolumn{1}{c}{kT} &
\multicolumn{1}{c}{\nh} &
\multicolumn{1}{c}{\eline} &
\multicolumn{1}{c}{\eqw} &
\multicolumn{1}{c}{$\chi^2$/dof} &
\multicolumn{1}{c}{Flux$^{(1)}$} \\

\multicolumn{1}{l|}{} &
\multicolumn{1}{c}{[eV]} &
\multicolumn{1}{c}{[\oexpo{20}cm$^{-2}$]} &
\multicolumn{1}{c|}{per instr.} &
\multicolumn{1}{c}{[eV]} &
\multicolumn{1}{c}{[\oexpo{20}cm$^{-2}$]} &
\multicolumn{1}{c}{[eV]} &
\multicolumn{1}{c}{[eV]} &
\multicolumn{1}{c}{per instr.} &
\multicolumn{1}{c}{[erg cm$^{-2}$ s$^{-1}$]} \\

\noalign{\smallskip}\hline\noalign{\smallskip}
AO1-pn   &  95.1       & 7.1 & 2138/309 & 85.8$\pm$0.5 & 4.1$\pm$0.1 & 290$\pm$5  & $-$148 & 589/307 & 3.65\expo{-12}\\
CAL-pn   &  =1$^{(2)}$ & =1  &          & =1	       & =1	     & =1	  & $-$148 &         & 3.44\expo{-12}\\
AO1-MOS1 &  =1         & 4.9 &  845/109	& =1	       & 2.9$\pm$0.2 & 302$\pm$2  & $-$159 & 156/108 & 3.33\expo{-12}\\
AO1-MOS2 &  =1         & 5.3 &          & =1	       & 3.2$\pm$0.2 & =3	  & $-$159 &         & 3.62\expo{-12}\\
AO1-RGS1 &  103        & 7.1 &  349/221	& 82.2$\pm$2.4 & 5.0$\pm$0.6 & =1         & --   & 318/221 & 2.62\expo{-12}\\
AO1-RGS2 &  109        & =5  &  	& 85.2$\pm$2.6 & =5          & =1	  & --	 &         & 2.50\expo{-12}\\
CAL-RGS1 &  112        & =5  &  	& 87.8$\pm$1.6 & =5          & =1         & --   &         & 2.66\expo{-12}\\
CAL-RGS2 &  108        & =5  &          & 87.6$\pm$1.5 & =5          & =1	  & --   &         & 2.62\expo{-12}\\
\noalign{\smallskip}\hline\noalign{\smallskip}
ACIS     &  81.1       & 19  & 253/38   & 87.8$\pm$1.0 & $<$1.6      & 290fixed   & --   & 91/38   & 3.34\expo{-12}\\
\noalign{\smallskip}\hline\noalign{\smallskip}
\end{tabular}

$^{(1)}$ Observed flux 0.1$-$2.4 keV; 
$^{(2)}$ ``=n" denotes fit parameter is linked with parameter in line n
\label{tab-fits}
\end{table*}

We also tried non-magnetic neutron star atmosphere models 
\citep[e.g.][]{2002A&A...386.1001G,2002nsps.conf..263Z}. Iron and solar mixture atmospheres
cause too many absorption features and deviations from a blackbody model in particular
at energies between 0.5 and 1.0 keV which are not seen in the measured spectra. On the other 
hand the spectrum of a pure hydrogen model is similar in shape to that of a blackbody and 
does not fit the data either. It also results in a much lower effective temperature of about 
30 eV which would predict a far too high optical flux \citep[see][]{1996ApJ...472L..33P}.
Using the models of \citet{2002A&A...386.1001G} $\chi^2_r$ values of
12, 29 and 12 (for the combined EPIC spectra) are obtained for Fe, solar mixture and H atmospheres, respectively.
A two-temperature blackbody fit results in $\chi^2_r$ = 6.6 with kT$_1$ = 33 eV and 
kT$_2$ = 94 eV. The strong residuals at energies below 500 eV remain. Such residuals are
not seen in fits to the spectra of \rxa\ with typical reduced $\chi^2_r$
values of 1.5 using the same version of spectral response files, clearly excluding 
calibration problems as origin.

An acceptable fit was found when including a broad absorption line in the model.
Because of the low line energy and the large line width the low resolution of the EPIC detectors
and the limited band pass of the RGS instruments do not allow to independently vary energy 
and width of the line. A broad line centered 
at the lowest energies reachable by the detectors ($\sim$120 eV) leaks only partially 
into the sensitive energy band and can therefore not be distinguished from a somewhat 
narrower line at somewhat higher energy. All we can say is that the absorption feature 
is extended up to $\sim$500 eV.
Models with an absorption line at 300 eV with $\sigma$ = 100 eV or a line at 200 eV
with $\sigma$ = 150 eV can not be distinguished. Also multiple lines at e.g. 150 eV and 300 eV
which are correspondingly narrower can not be excluded. However, too broad lines are probably 
unrealistic \citep[see e.g.][ if interpreted as cyclotron resonance]{2001ApJ...560..384Z} 
and we therefore use in our fits a single absorption line of Gaussian shape 
with fixed $\sigma$ = 100 eV. This yields a best fit with $\chi^2_r$ = 1.67 for 636 dof 
(Fig.~\ref{fig-spectra})
with parameters listed in Table~\ref{tab-fits}.
The line energy was allowed to be fit individually for the two EPIC instruments but is 
consistent within the errors. Because the sensitive energy band of the RGS instruments
covers the broad line only partially, we forced all line parameters to be the same in the fit 
for the pn and all RGS spectra simultaneously. For the same reason we do not give values for the line 
equivalent width (EQW) for the RGS spectra. The $\chi^2$ obtained from the RGS spectra for 
this model is reduced by 31 (without adding additional dofs) as compared to the blackbody model
which may be accepted for RGS if treated by its own.

\begin{figure}
\begin{center}
\resizebox{8.6cm}{!}{\includegraphics[clip=,angle=-90]{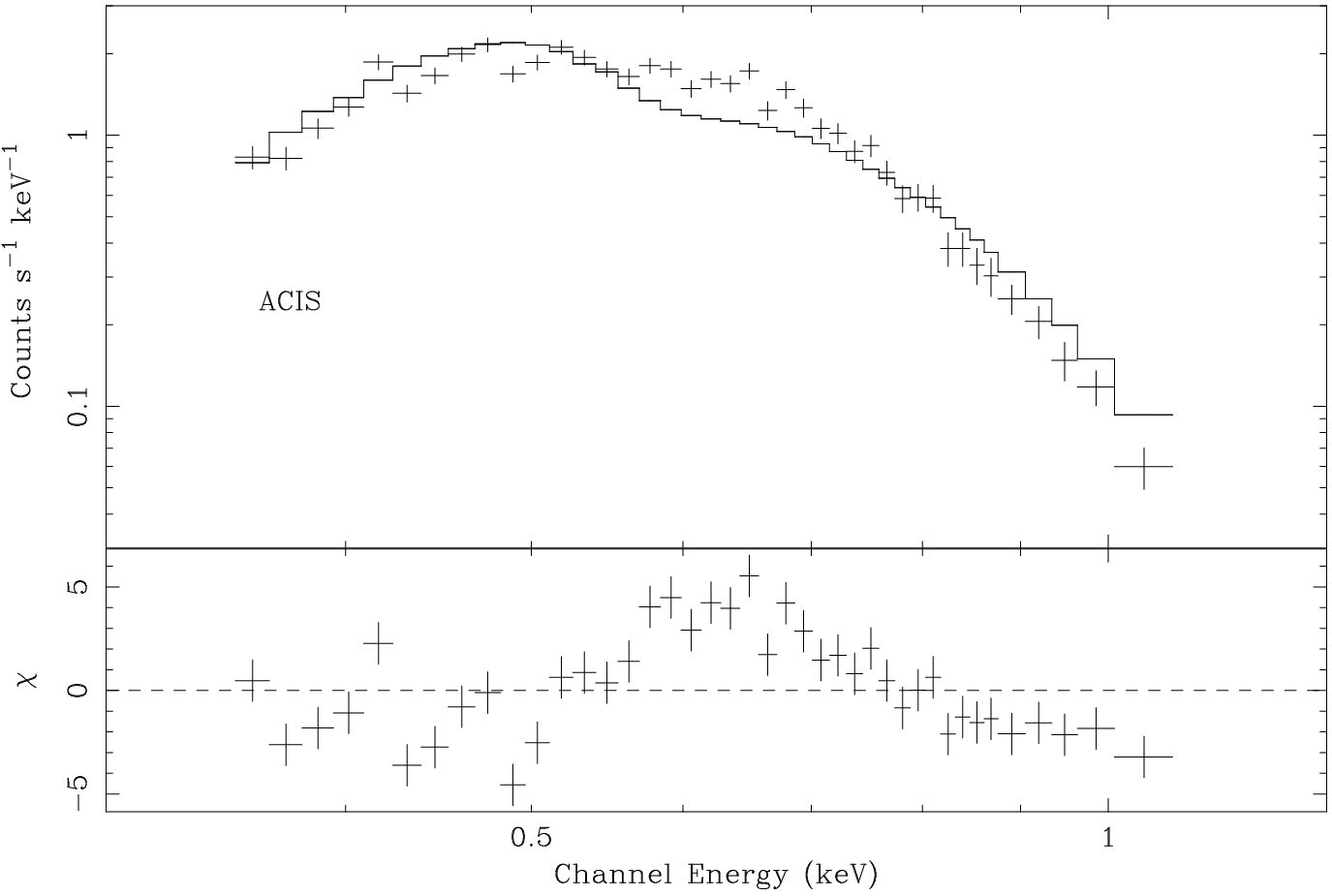}}
\resizebox{8.6cm}{!}{\includegraphics[clip=,angle=-90]{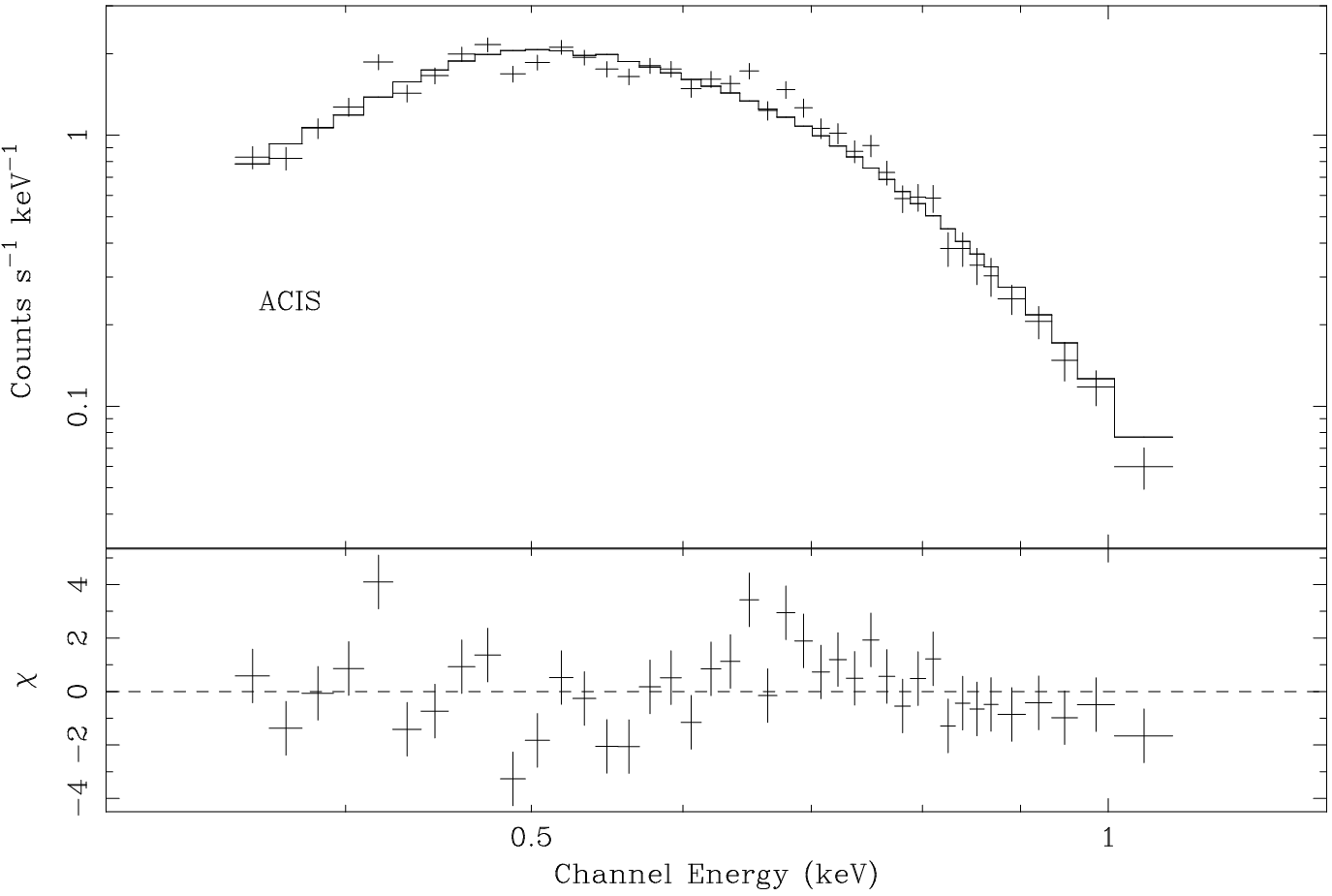}}
\end{center}
\caption{Blackbody model fits with (bottom) and without absorption line to the ACIS-S spectrum of \rbs.}
\label{fig-acis}
\end{figure}

We re-analyzed the ACIS-S spectrum of \rbs, published by \citet{2002A&A...381...98H}, using
the re-processed data available in the Chandra archive and new calibration which allows
a spectral analysis down to 350 eV. We fit blackbody with and without absorption line
to the spectrum, with all line parameters fixed at the values derived from the fit to the
XMM-Newton spectra. The fits are shown in Fig.~\ref{fig-acis} and the model parameters
reported in Table~\ref{tab-fits}. The results are in good agreement to those obtained from the 
XMM-Newton data.

\section{X-ray pulsations}

\begin{figure}
\begin{center}
\resizebox{8.6cm}{!}{\includegraphics[clip=,angle=-90]{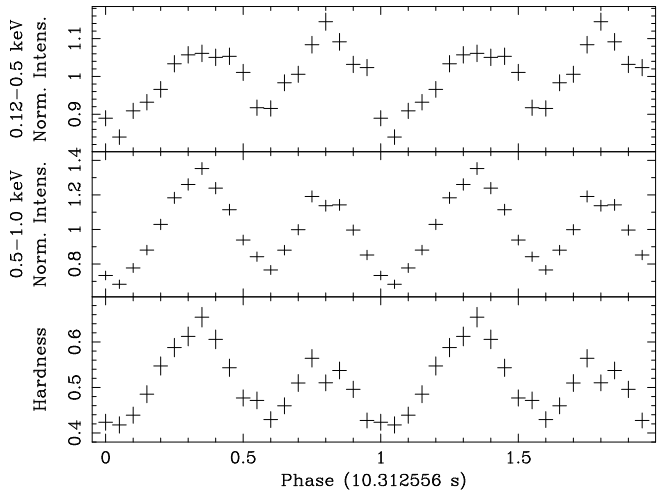}}
\end{center}
\caption{Pulse profile of \rbs\ in the 0.12$-$0.5 keV (soft) and 0.5$-$1.0 keV (hard) 
         energy bands, together with the ratio hard/soft, obtained
	 from the EPIC-pn data of the Jan. 2003 observation.}
\label{fig-pulse}
\end{figure}

To investigate spectral changes with pulse phase, light curves in the energy bands 
0.12$-$0.5 and 0.5$-$1.0 keV were 
folded with a pulse period of 10.31 s. This long period with double peaked profile
is suggested by the presence of a second peak in the power spectrum 
\citep{haberl2002COSPAR}. The different relative strength of the two peaks in the two
energy bands and the related significant difference in hardness ratio (Fig.~\ref{fig-pulse})
strongly supports a 10.31 s spin period of \rbs, twice as long as originally thought.

\section{Discussion}

Spectral analysis of the INS \rbs\ using data obtained by the X-ray
instruments aboard XMM-Newton revealed significant deviations from a Planckian energy 
distribution at energies below $\sim$500 eV. Also, non-magnetic atmosphere models
yield unacceptable fits. We find that a model including a broad 
absorption line on top of the blackbody continuum can reproduce the observed spectra.
For a Gaussian line with fixed $\sigma$=100 eV we derive an energy of the line of 
$\sim$300 eV and an equivalent width of $-$150 eV. However, the spectral resolution 
of the EPIC instruments at these energies and the restricted energy coverage of the RGS
are not sufficient to exclude line energies
as low as 100 eV or multiple lines which are unresolved.

Cyclotron resonance absorption features in the 0.1$-$1.0 keV band are expected in spectra from 
magnetized neutron stars with field strengths in the range of \oexpo{10}$-$\oexpo{11} G
or 2\expo{13}$-$2\expo{14} G if caused by electrons or protons, respectively 
\citep[see e.g.][]{2001ApJ...560..384Z,2002nsps.conf..263Z}. 
A strong magnetic field for \rbs\ of $\sim$2\expo{14} G was first derived by
\citet{2002A&A...381...98H} from a first estimate of the spin period derivative based on 
a ROSAT and a Chandra observation. The first of the two XMM-Newton observations described 
here, however, put this into question \citep[for preliminary results see][]{haberl2002COSPAR} and placed
only an upper limit of \oexpo{14}~G. A detailed temporal analysis of the two XMM-Newton 
observations and a re-analysis of the Chandra data will be published in an
accompanying paper (Hambaryan et al. in preparation).

Recently, several radio pulsars with magnetic field strength of 
a few \oexpo{13} G with similar long spin period as \rxb\ and \rbs\ were 
discovered \citep{2000ApJ...541..367C,2002MNRAS.335..275M}. 
If one assumes a comparable field strength for \rbs\ which may be justified
if the ROSAT discovered radio-quiet INSs are cases where the radio beam does not cross the Earth
\citep{2002ApJ...570L..79K,motch}, but are otherwise similar objects, one would
favor cyclotron resonance absorption due to protons to be responsible for the
features seen in the spectra of \rbs. Then our measured line energies between 
100 and 300 eV translate into magnetic field strengths of 2$-$6\expo{13} G for a 
neutron star with canonical mass (1.4 M$_{\sun}$) and radius (10 km). 

By folding the EPIC-pn data of \rbs\ in two energy bands with a period of 10.31 s
we see hardness ratio changes with pulse phase. The two
pulses show different hardness which may be explained by variations in cyclotron 
absorption between two magnetic poles. 
Detailed phase resolved spectroscopy which exceeds the scope of this 
letter, is in progress and will be reported elsewhere.

Variations in the hardness ratio with pulse phase were also reported from \rxb.
Formal fits using a blackbody spectrum attenuated by photo-electric absorption
performed by \citet{2001A&A...365L.302C} to the pulse-phase resolved EPIC-pn spectra around
hardness maximum and minimum yielded changes in column density, but little variations in
blackbody temperature. This could be explained in terms of energy-dependent beaming effects
with softer photons more strongly beamed than harder photons. From their polar cap modeling 
of the spin pulse profile the authors conclude that the derived geometries are at best marginally
compatible with surface temperature variation but more satisfactory explained by an 
accretion model. However, the possibility of accretion of interstellar matter is now most 
likely excluded by the measured high proper motion of \rxb\ \citep{motch}. 
On the other hand, the variations with pulse phase could be naturally explained by 
cyclotron absorption, because we may view surface areas with different magnetic field 
strength and orientation during the rotation of the neutron star. A cyclotron absorption 
line at the low energy end of the sensitive
energy band of the EPIC instruments would cause deviations from a Planckian shape 
at these energies
which, so far, were not recognized due to lack of statistics and/or energy resolution
and preliminary calibration of the new generation instruments.

\citet{2001ApJ...560..384Z} calculate spectra emerging from ultra-magnetized 
neutron stars covered by non-degenerate, pure hydrogen plasma. 
They explored ranges in magnetic field strength from \oexpo{13} G to \oexpo{15} G
and in luminosity from \oergs{34} to \oergs{36}. Although the luminosity of \rbs\
may only be around \oergs{31-32} their results for the line EQW shows little dependence
on luminosity and may be compared to the value of 150 eV we find for \rbs.
Their predictions for cyclotron line widths and equivalent width and the overall blackbody-like
continuum for the case of a dipolar magnetic field (which broadens the line due to
B field variations in magnitude and direction along the neutron star surface)
are consistent with our results for
\rbs. This suggests, that at least for \rbs\ and probably also \rxb\ a H atmosphere
exists on these neutron stars and argues against a condensed matter surface as it was
proposed to explain the combined spectral energy distribution in the optical and
X-ray bands \citep{burwitz,zane03}. However, an optically thick H atmosphere produces
too much optical flux \citep{1996ApJ...472L..33P}, a problem which may be reduced
by thin atmosphere models as was applied by \citet{motch} to \rxb.

\begin{acknowledgements}
The XMM-Newton project is supported by the Bundesministerium f\"ur Bildung und
For\-schung / Deutsches Zentrum f\"ur Luft- und Raumfahrt (BMBF / DLR), the
Max-Planck-Gesellschaft and the Heidenhain-Stif\-tung. 
\end{acknowledgements}

\bibliographystyle{apj}
\bibliography{ins,general,myrefereed,myunrefereed,mytechnical}

\end{document}